\documentclass[%
 reprint,
 amsmath,amssymb,
 aps,prc
]{revtex4-2}
\usepackage[colorlinks, citecolor=red]{hyperref}
\usepackage{graphicx}
\usepackage{dcolumn}
\usepackage{bm}
\usepackage{natbib}
\usepackage{multirow}
\begin{document}
\preprint{APS/123-QED}
\title{Effects of \texorpdfstring{$\phi$}{} and \texorpdfstring{$\sigma^{*}$}{}-meson on properties of hyperon stars including \texorpdfstring{$\Delta$}{} resonance}

\author{Chen Wu$^{1}$\footnote{Electronic address: wuchenoffd@gmail.com} and Wenjun Guo$^{2}$ } \affiliation{
\small 1. Xingzhi College, Zhejiang Normal University, Jinhua, 321004, Zhejiang, China\\
\small 2. College of Science, University of Shanghai for Science and Technology,
Shanghai, 200093, China.}

\begin{abstract}
In this work, we research  the properties of neutron stars using the nonlinear Relativistic Mean-Field (RMF) theory and consider multiple degrees of freedom inside neutron stars, including hyperons and $\Delta$ resonances. We investigate different coupling parameters $x_{\sigma \Delta}$ between $\Delta$ resonances and nucleons and compare the differences between neutron stars with and without strange mesons $\sigma^*$ and $\phi$. These effects include particle number distributions, equations of state (EOS), mass-radius relations, and tidal deformabilities. To overcome the "hyperon puzzle," we employ the $\sigma-cut$ scheme to obtain neutron stars with masses up to $2M_{\odot}$. We find that strange mesons appear at around 3$\rho_0$ and reduce the critical density of baryons in the high density region. With increasing coupling parameter $x_{\sigma \Delta}$, the $\Delta$ resonances suppress hyperons, leading to a shift of the critical density towards lower values. The early occurrence of $\Delta$ resonances may play a crucial role in the stability of neutron stars. Strange mesons soften the EOS slightly, while $\Delta$ resonances predominantly soften the EOS in the low density region. By calculating tidal deformabilities and comparing with astronomical event GW170817, we find that the inclusion of $\Delta$ resonances decreases the radius of neutron stars.
\end{abstract}

\maketitle

\section{\label{sec:level1}INTRODUCTION}

Observations of massive neutron stars, such as PSR J1614-2230 with a mass of $1.908\pm0.016M_{\odot}$, have provided important constraints on the EOS of high density nuclear matter \cite{Demorest:2010bx,NANOGrav:2017wvv,Fonseca:2016tux,Ozel:2010bz}. Similarly, PSR J0348+0432 ($2.01 \pm0.04$ times the solar mass $M_{\odot}$) and MSP J0740+6620 with a mass of $2.08^{+0.07}_{-0.07}M_{\odot}$ \cite{fonseca2021refined,NANOGrav:2019jur}   have also contributed to constraining the EOS (from NICER observations \cite{riley2021nicer}). The first multi-messenger gravitational wave event GW170817 observed by LIGO-Virgo Collaboration (LVC) set constraints on the tidal deformability \cite{LIGOScientific:2018hze,Coughlin:2018miv} of the involved stars. The compactness-radius relation predicts a radius of $12\leq R_{1.4} \leq 13$ km for a standard mass neutron star with $M=1.4M_{\odot}$. These astronomical observations not only constrain the tidal deformability of a 1.4 $M_{\odot}$ neutron star but also shed light on the strong interactions in nuclear matter at high density.

\par{}

Due to the high core density inside neutron stars, as the nucleon density increases, the strong interaction among hadrons leads to the excitation of hyperons \cite{Schaffner:1995th,Wu:2011zzb,Burgio:2021vgk,Logoteta:2021iuy}, kaon meson condensation \cite{Thapa:2020usm,Maruyama:2005tb,Brown:2005yx,Shao:2010zz,Char:2014cja}, or $\Delta$ resonances \cite{Sun:2018tmw,Maslov:2017lkx,Kolomeitsev:2016ptu,Sedrakian:2022kgj,Drago:2014oja,Glendenning:1984jr,Dexheimer:2021sxs,Marquez:2022fzh,Schurhoff_2010} in the neutron star interior. These forms of matter have significant impacts on the structure and evolution of the stars. While the presence of hyperons in dense nuclear matter is unavoidable, their appearance results in a significant softening of the EOS, leading to a reduction in the maximum mass of neutron stars, commonly referred to as the hyperon puzzle \cite{burgio2011hyperon,Lonardoni:2014bwa,Bombaci:2016xzl}. Despite various speculations about the appearance of hyperons in dense nuclear matter, discussions about $\Delta$ resonances have been limited. This is because the coupling parameters of $\Delta$ with nucleons are not well determined, either experimentally or theoretically, and different coupling parameters can have a significant impact on the final results \cite{Glendenning:1984jr, Glendenning:1991es}. Additionally, the appearance of $\Delta$ resonances causes the EOS of neutron star matter to soften, giving rise to a $\Delta$ puzzle, similar to the hyperon puzzle found in some literature \cite{Drago:2014oja}. To ensure the stiffness of the equation of state and the existence of massive neutron stars, researchers have employed relativistic mean field theory with the inclusion of the sigma-cut scheme \cite{Maslov:2015lma} and density covariant functional theory to study neutron stars containing hyperons \cite{bonanno2012composition, Colucci:2013pya, Fortin:2016hny, Chen:2007kxa, Drago:2014oja, Li:2019fqe, Li:2019fqe, li2019implications}.

\par{}

The resolution of the hyperon puzzle requires additional repulsive interactions between baryons \cite{Vidana:2013nxa} to counteract this attractive mechanism. These repulsive interactions include:
(a) Increasing the repulsive hyperon three-body force \cite{Lonardoni:2014bwa}.
(b) Deconfinement phase transition to quark matter below the hyperon threshold \cite{Klahn:2013kga}.
(c) Enhancing the repulsive hyperonic interactions through the exchange of vector mesons \cite{Bhuyan:2016pgk}. The relativistic mean field theory is an effective field theory used to handle the interactions between hadrons (nucleons) in a relativistic framework. The relevant degrees of freedom in this theory are baryons interacting through the exchange of $\sigma$, $\omega$, and $\rho$ mesons. The scalar meson $\sigma$ provides intermediate-range attraction, the vector meson $\omega$ provides short-range repulsion, and the vector-isovector meson $\rho$ describes the difference between neutrons and protons. Over the years, the $\sigma$, $\omega$, and $\rho$ mesons have been widely applied in various aspects of neutron star research. However, there should also be scalar meson $\sigma^*$ and vector meson $\phi$ involved in the interactions between hyperons. These two mesons specifically interact between hyperons and do not participate in interactions between nucleons. In this work, we employ an alternative approach to overcome the softening of the equation of state  caused by multiple degrees of freedom, namely the ${\sigma}$-cut scheme \cite{Maslov:2015lma}. This scheme suggests that in high density region, a considerable decrease in the strength of the ${\sigma}$ meson  reduces the decrease in the effective mass of nucleons, leading to an increase in the particle chemical potential and ultimately resulting in the stiffening of the EOS \cite{Ma:2022fmu, Wu:2020yqp}. In this article, we use the IUFSU model \cite{Fattoyev:2010mx,Grill:2014aea} to research neutron star matter including hyperons, ${\Delta}$ resonance, and strange mesons ($\sigma^*, \phi$) within the framework of the ${\sigma}$-cut scheme.
\par
This article is organized as follows. The theoretical framework is presented in Section 2. Then we will consider the effects of strange mesons ($\sigma^*, \phi$) and ${\Delta}$ resonance within  the framework of the $\sigma$-cut scheme. Finally, some conclusions are presented.

\section{\label{sec:level2}theoretical framework}

In this section, we introduce the IUFSU model to consider the properties of the neutron star (NS) including hyperons and ${\Delta}$ resonance, the Lagrangian density of neutron star matter is given by:
\begin{widetext}
\begin{equation}
\begin{aligned}
\mathcal{L}=&\sum_{B}\bar{\psi}_{B}[\emph{$i$}\gamma^{\mu}\partial{\mu}-m_B+g_{\sigma B}\sigma + g_{\sigma^{*} B}\sigma^{*} - g_{\omega B}\gamma^{\mu}\omega_{\mu}-g_{\phi B}\gamma^{\mu}\phi_{\mu}-{g_{\rho B}}\gamma^{\mu}\vec{\tau}\cdot\vec{\rho^{\mu}}]\psi_{B}+\\
&\sum_{D}\bar{\psi}_{D}[\emph{$i$}\gamma^{\mu}\partial{\mu}-m_D+g_{\sigma D}\sigma-g_{\omega D}\gamma^{\mu}\omega_{\mu}-{g_{\rho D}}\gamma^{\mu}\vec{\tau}\cdot\vec{\rho^{\mu}}]\psi_{D}+\\
&\frac{1}{2}\partial_{\mu}\sigma\partial^{\mu}\sigma-\frac{1}{2}m_{\sigma}^{2}\sigma^{2}-\frac{\kappa}{3!}(g_{\sigma N}\sigma)^3-\frac{\lambda}{4!}(g_{\sigma N}\sigma)^{4}-\frac{1}{4}F_{\mu \nu}F^{\mu \nu}+\frac{1}{2}m_{\omega}^{2}{\omega_{\mu}}\omega^{\mu}-\\
&\frac{1}{2}(\partial_{\mu}\sigma^*\partial^{\mu}\sigma^*-m_{\sigma^*}^{2}\sigma^{*2})
+\frac{\xi}{4!}(g_{\omega N}^{2}\omega_{\mu}\omega^{\mu})^{2}+\frac{1}{2}m_{\rho}^{2}\vec{\rho}_{\mu}\cdot\vec{\rho}^{\mu}-\frac{1}{4}\vec{G}_{\mu \nu}\vec{G}^{\mu \nu}+\\
&\frac{1}{4}{\Phi}_{\mu \nu}{\Phi}^{\mu \nu}+\frac{1}{2}m_{\phi}^{2}{\phi_{\mu}}\phi^{\mu}
+\Lambda_{\nu}(g_{\rho N}^{2}\vec{\rho}_{\mu}\cdot\vec{\rho}^{\mu})(g_{\omega N}^{2}\omega_{\mu}\omega^{\mu})+\sum_{l} \bar{\psi}_{l}[i{\gamma}^{\mu}\partial{\mu}-m_{l}]{\psi}_{l},
\label{eq:one}
\end{aligned}
\end{equation}
\end{widetext}

with the field tensors
\begin{equation}
\begin{aligned}
&\emph{F}_{\mu \nu}=\partial_{\mu}\omega_{\nu}-\partial_{\nu}\omega_{\mu}\\
&\vec{G}_{\mu \nu}=\partial_{\mu}\vec{\rho}_{\nu}-\partial_{\nu}\vec{\rho}_{\mu}\\
&{\Phi_{\mu \nu}}=\partial_{\mu}\phi_\nu-\partial_{\nu}\phi_{\mu},
\end{aligned}
\end{equation}

This model includes the baryon octet and two leptons ($p, n, e^-, \mu^-, \Lambda^{0}, \Xi^{0},\Xi^{-}, \Sigma^{+}, \Sigma^{0}, \Sigma^{-}$) as well as $\Delta$ resonances ($\Delta^{++}, \Delta^{+}, \Delta^{0}, \Delta^{-}$). The strong interactions between baryons are mediated by the isoscalar-scalar mesons $\sigma$, $\sigma^*$, isoscalar-vector mesons $\omega$, $\phi$ and isovector-vector meson $\rho$, each with their respective masses and coupling constants. The  $\vec{\tau}$ represent the isospin operator for the isovector-vector meson fields. The parameter $\Lambda_{\nu}$ is used to modify the density dependence of the symmetry energy. The self-interactions of the isoscalar mesons (through the $\kappa$, $\lambda$, and $\xi$ terms) are necessary to obtain an appropriate equation of state for symmetric nuclear matter.  The several parameters of the IUFSU model are listed in Table \ref{tab:Table 2.}.

\par{}
Finally, by means of the Euler-Lagrange equation, the
equations of motion for baryons and mesons are given by:
\begin{widetext}
\begin{equation}
\begin{aligned}
&m_{\sigma}^{2}\sigma+\frac{1}{2}{\kappa}g_{\sigma N}^{3}{\sigma}^{2}+\frac{1}{6}{\lambda}g_{\sigma N}^{4}{\sigma}^{3}=\sum_{B}g_{\sigma B}{\rho}_{B}^{S}+\sum_{D}g_{\sigma D}{\rho}_{D}^{S}\\
&m_{\omega}^{2}{\omega}+\frac{\xi}{6}g_{\omega N}^{4}{\omega}^{3}+2{\Lambda_{\nu}}g_{\rho N}^{2}g_{\omega N}^{2}{\rho}^{2}{\omega}=\sum_{B}g_{\omega B}{\rho}_{B}+\sum_{D}g_{\omega D}{\rho}_{D}\\
&m_{\rho}^{2}{\rho}+2{\Lambda}_{\nu}g_{\rho N}^{2}g_{\omega N}^{2}{\omega}^{2}{\rho}=\sum_{B}{g_{\rho B}}{\tau}_{3B}{\rho}_{B}+\sum_{D}{g_{\rho D}}{\tau}_{3D}{\rho}_{D}\\
&m_{\phi}^{2}{\phi}=\sum_{B}g_{\phi B}{\rho}_{B}\\
&m_{\sigma^*}^{2}{\sigma^*}=\sum_{B}g_{\sigma^* B}{\rho}_{B}^{S},
\end{aligned}
\end{equation}
\end{widetext}
where $\rho_{B(D)}$ and $\rho_{B(D)}^{S}$ are the baryon ($\Delta$) density and the scalar density, which are given by:
\begin{equation}
\begin{aligned}
&\rho_B = \gamma \frac{ k_{fB}}{6\pi^2}\\
&\rho_B^S = \gamma \frac{ M^{*}}{4\pi^2}[-M^{*2}\ln(\frac{k_{fB}+E^{*}_{fB}}{M^{*2}}) + k_{fB}E^{*}_{fB}]
\end{aligned}
\end{equation}
$\gamma=2$ for baryons and $\gamma=4$ for $\Delta$ resonance. Here $E^{*}_{fB}=\sqrt{M^{*^2} + k_{fB}^2}$.
\par{}
The next step is to fit the coupling parameters between  baryons (nucleons, hyperons and $\Delta$) and meson fields.
\begin{table*}
\caption{\label{tab:Table 2.}
Coupling constants and nonlinear interacting parameters for the IUFSU model and the meson masses $M_{\sigma}=491.5MeV$, $M_{\omega}=786MeV$, $M_{\rho}=763MeV$.
}
\begin{ruledtabular}
\begin{tabular}{lccccccr}
\textrm{Model}&
\textrm{$g_{\sigma}$} &
\textrm{$g_{\omega}$} &
\textrm{$g_{\rho}$} &
\textrm{$\Lambda_{\nu}$}&
\textrm{$\kappa$} &
\textrm{$\lambda$} &
\textrm{$\xi$} \\
IUFSU & 9.9713 & 13.0321 & 13.5899 & 0.046 &3.37685 & 0.000268 & 0.03  \\
\end{tabular}
\end{ruledtabular}
\end{table*}

\par
For the vector meson-hyperon couplings, we use the SU(6) symmetry to obtain the vector couplings constants:
\begin{equation}
\begin{aligned}
&g_{\omega \Lambda}=g_{\omega \Sigma}=2g_{\omega \Xi}=\frac{2}{3}g_{\omega N},\\
&g_{\rho \Lambda}=0, g_{\rho \Sigma}=2g_{\rho \Xi}=2g_{\rho N},\\
&2g_{\phi \Lambda}=2g_{\phi \Sigma}=g_{\phi \Xi}=\frac{-2\sqrt{2}}{3}g_{\omega N}.
\end{aligned}
\end{equation}

The scalar couplings are typically determined by fitting hyperon potentials with $U_{Y}^{(N)}=g_{\omega Y}\omega_{0}-g_{\sigma Y}\sigma_{0}$, where $\omega_{0}$ and  $\sigma_{0}$  are the values of the  vector and  scalar meson strengths at nuclear saturation density \cite{Schaffner:1993qj}. The hyperon-nucleon potentials of $\Lambda$, $\Sigma$, and $\Xi$  are chosen as $U^N_{\Lambda}=-30$ MeV, $U^N_{\Sigma}=30$ MeV, and $U^N_{\Xi}=-18$ MeV \cite{Hu:2021ket,Fortin:2017cvt,AGSE885:1999erv}. Table \ref{tab:Table 3.} lists the calculated values of the meson-hyperon couplings, where $x_{\sigma Y}={g_{\sigma Y}}/{g_{\sigma N}}$.

While strange mesons do not interact with nucleons and resonances, so the corresponding coupling constants are zero. The masses of the strange mesons $\phi$ and $\sigma^*$ are $M_{\phi}=1020$MeV and $M_{\sigma^*}=975$MeV, respectively.

\begin{table}.
\caption{\label{tab:Table 3.}
scalar meson hyperon coupling constants for the IUFSU model discussed in the text.
}
\begin{ruledtabular}
\begin{tabular}{lccc}
\textrm{} &
\textrm{$\Sigma$} &
\textrm{$\Lambda$} &
\textrm{$\Xi$} \\
$x_{\sigma Y}= g_{\sigma Y}/g_{\sigma Y}$  & 0.45219 & 0.615796 & 0.305171
\end{tabular}
\end{ruledtabular}
\end{table}
\par
For the scalar meson $\sigma^*$, we treat its coupling purely phenomenologically so as to satisfy the potential depths $U_{\Sigma}^{(\Xi)}$ $\simeq$ $U_{\Lambda}^{(\Xi)}$ $\simeq$ $U_{\Xi}^{(\Xi)}$ $\simeq$ $U_{\Lambda}^{(\Lambda)}$ $\simeq$ $2U_{\Sigma}^{(\Lambda)} = 40MeV$. This yield $g_{\sigma^*\Lambda}/g_\sigma=g_{\sigma^*\Sigma}/g_\sigma=0.69$, $g_{\sigma^*\Xi}/g_\sigma=1.25$ \cite{Schaffner:1993nn}.
\par

Due to the scarcity of experimental data and theoretical calculations regarding the $\Delta$ resonance, there is uncertainty in the coupling parameters between the $\Delta$ resonances and meson fields($\sigma, \omega, \rho$). Therefore, we use the method explored in the literature \cite{Li:1997yh, Jin:1994vw}, which assume that the scalar coupling ratio $x_{\sigma \Delta}=g_{\sigma \Delta}/g_{\sigma N} > 1$ and choose a value close to the mass ratio of the $\Delta$ and the nucleon \cite{Kosov:1998gp}. In this work, we use three different values for $x_{\sigma \Delta}$ ($x_{\sigma \Delta}=1.05$, 1.1, and 1.15) \cite{Drago:2013fsa}. For $x_{\omega \Delta}$ and $x_{\rho \Delta}$, we take $x_{\omega \Delta}=g_{\omega \Delta}/g_{\omega N}=1.1$ and $x_{\rho \Delta}=g_{\rho \Delta}/g_{\rho N}=1$ as the literature \cite{Thapa:2021kfo}.

\par{}
When neutrinos are not captured, the set of equilibrium chemical potential relations under general conditions is as follows::
\begin{equation}
\mu_{i} = \mu_{n} - q_i\mu_{e^-}
\end{equation}
where $q_{i}$ is the charge of $i-$th baryon, and the charge neutrality condition is fulfilled by:
\begin{equation}
\sum_{B}q_B\rho_B+\sum_{D}q_D\rho_D = \rho_{e^-} + \rho_{\mu^-}.
\end{equation}
The chemical potentials of baryons, $\Delta$ and leptons are given by:
\begin{equation}
\mu_{i}=\sqrt{k_{F}^{i 2}+m_{i}^{* 2}}+g_{\omega i}\omega+g_{\rho i}\tau_{3 i} \rho +g_{\phi i}\phi , i=B, D
\end{equation}
\begin{equation}
\mu_{l}=\sqrt{k_{F}^{l 2}+m_{l}^{2}}, l= e^-, \mu^-
\end{equation}
where $k_{F}^{i}$ is the Fermi momentum and the $m_{i}^{*}$ is the effective mass of baryon and $\Delta$ resonances, which is computed via $m_{i}^{*}=m_{i}-g_{\sigma i}\sigma-g_{\sigma^* i}\sigma^*$, and $k_{F}^{l}$ is the Fermi momentum of the lepton $l$($\mu^-,e^-$).
\par
The total energy density can be given as
\begin{equation}
\begin{aligned}
\varepsilon=&\sum_{i=B,D}\frac{\gamma}{(2\pi)^{3}}\int_{0}^{k_{F}^i}\sqrt{m_{i}^{*}+k^{2}}d^{3}k+\frac{1}{2}m_{\omega}^{2}\omega^{2}\\
&+\frac{\xi}{8}g_{\omega N}^{4}\omega^{4}+\frac{1}{2}m_{\sigma}^{2}\sigma^{2}+\frac{\kappa}{6}g_{\sigma N}^{3}\sigma^{3}+\frac{\lambda}{24}g_{\sigma N}^{4}\sigma^{4}\\
&+ \frac{1}{2}m_{\phi}^{2}\phi^{2} + 3\Lambda_{\nu}g_{\rho N}^{2}g_{\omega N}^{2}\omega^{2}\rho^{2} + \frac{1}{2}m_{\rho}^{2}\rho^{2}\\
&+\frac{1}{2}m_{\sigma^{*}}^{2} \sigma^{2*}+\frac{1}{\pi^{2}}\sum_{l}\int_{0}^{k_{F}^l}\sqrt{k^{2}+m_{l}^{2}}k^{2}dk,
\label{eq6}
\end{aligned}
\end{equation}
And the expression of pressure reads
\begin{equation}
P = \sum_{i=B,D}\rho_{i}\mu_{i}+\sum_{l=\mu^-,e^-}\rho_{l}\mu_{l} - \varepsilon,
\end{equation}
Once the equation of state is specified, the mass-radius relation and other relevant quantities of neutron star can be calculated by solving the Tolman-Oppenheimer-Volkoff (TOV) equation \cite{Baldo:1997ag}.
\begin{equation}
\begin{aligned}
\frac{dP(r)}{dr}=&-\frac{GM(r)}{r^{2}}\varepsilon(1+\frac{4\pi r^{3}P}{M(r)C^{2}})(1+\frac{P}{\varepsilon C^{2}})\\
&\times(1-\frac{2GM(r)}{rC^{2}})^{-1},
\end{aligned}
\end{equation}
\begin{equation}
dM(r)=4\pi r^{2}\varepsilon(r)dr
\end{equation}
The tidal deformability of $\Lambda$  related to the $l=2$ dimensionless tidal Love number $k_2$, is given by \cite{hinderer2008tidal,Postnikov:2010yn}.
\begin{equation}
\Lambda=\frac{2}{3}k_{2}{C_1}^{-5}
\end{equation}
where $C_1=GM/R$, the  $k_{2}$ can be fixed simultaneously with the structures of compact stars \cite{Hinderer:2009ca}.
\par
The $\sigma$-cut scheme \cite{Maslov:2015lma} involves introducing an additional sigma self-interaction term in the model \cite{Maslov:2015lma,Song:2011jh,Kolomeitsev:2015qia}, which results in a stiffer equation of state (EOS):
\begin{equation}
\Delta U(\sigma)=\alpha \ln(1+exp[\beta(f-f_{s,core})]).
\end{equation}

In this context, $f=g_{\sigma N}{\sigma}/M_{N}$ and $f_{s,core}=f_{0}+c_{\sigma}(1-f_{0})$. Here, $M_{N}$ represents the nucleon mass, and $f_{0}$ is the value of $f$ at saturation density, which equals 0.31 in the IUFSU model. $c_{\sigma}$ is a positive parameter that we can adjust.  In our previous work, we extensively discussed the choice of the parameter $c_{\sigma}$ \cite{Ma:2022fmu}. In this work, we adopt $c_{\sigma}=0.15$ to satisfy the constraint on the maximum mass. The constants $\alpha$ and $\beta$ have values of $4.822 \times 10^{-4}M_{N}^{4}$ and 120, respectively, following the settings in Ref. \cite{Maslov:2015lma}.
\par
\section{\label{sec:level3}Results}
\begin{figure}
\includegraphics{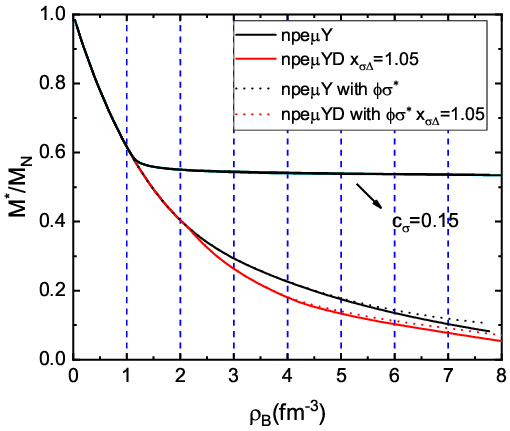}
\caption{Effective mass of nucleons versus baryon density in NS matter using and not using $\sigma$-cut scheme with considering $\sigma^*$ and $\phi$ or not.}
\label{fig1}
\end{figure}

First, we considered the effect of the $\sigma$-cut scheme on the IUFSU model. In Fig. \ref{fig1}, we display the ratio of the effective mass to the rest mass of nucleons as function of the baryon density. The dashed and solid lines represent the cases with and without strange mesons $\sigma^*$ and $\phi$, respectively. Here, $\rho_{0}$ is the saturation density, and we chose $x_{\sigma \Delta}=1.05$ to consider the $\Delta$ resonance. We can observe that when $\rho \leq \rho_{0}$, the effective mass is almost the same as in nucleons-only matter and remains unchanged by the $\sigma$-cut scheme. This suggests that the scheme does not alter the properties of nuclear matter at saturation, which is crucial. The inclusion of strange mesons $\sigma^*$ and $\phi$ slightly slows down the drop in effective mass $M^*$, but this effect disappears when the $\sigma$-cut scheme is applied. However, when $\rho$\textgreater$\rho_{0}$, the effective mass drops to around $0.55 M_{N}$, significantly suppressing the strength of the $\sigma$ meson field strength. This is precisely the desired effect achieved by using the $\sigma$-cut scheme.
\begin{figure}
\centering
\includegraphics{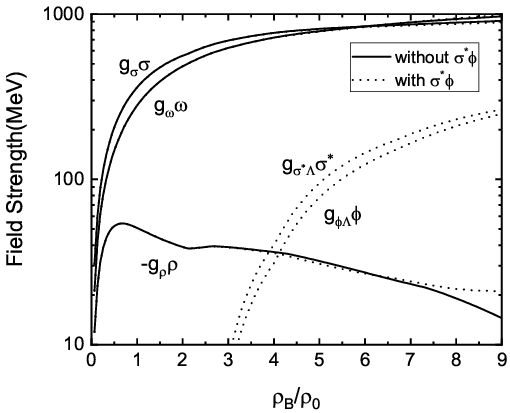}
\caption{Field strength of various mesons with considering $\sigma^*$ and $\phi$ or not.}
\label{fig2}
\end{figure}
\par
\begin{figure}
\centering
\includegraphics{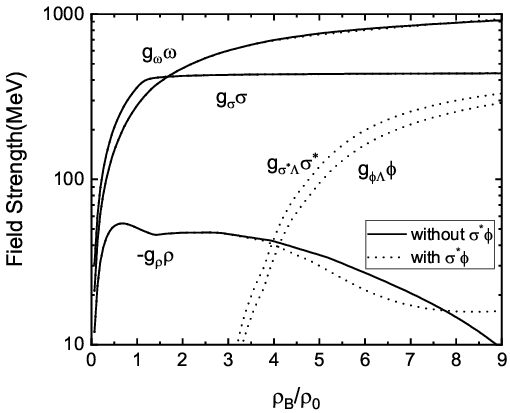}
\caption{Field strength of various mesons with considering $\sigma^*$ and $\phi$ or not, $c_{\sigma}$=0.15}
\label{fig3}
\end{figure}

The field strength of various mesons are shown in Fig. \ref{fig2}. When the baryon density is approximately 3$\rho_0$, the $\sigma^*$ and $\phi$ mesons emerge, while the field strength of $\sigma$ and $\omega$ mesons remains nearly unchanged. However, the field strength of the $\rho$ meson increases in the high-density region($6.6\rho_0$). And we also distinctly see from Fig. \ref{fig2}, the field strength of $\sigma^*$ is larger than $\phi$ and both increase with the baryon density. When we choose $\sigma$-cut scheme(Fig. \ref{fig3}), the field strength of $\sigma$ meson is truncated, and the field strength of the $\rho$ meson decreases around $4\rho_0$ \texttt{\textasciitilde} $6\rho_0$, and subsequently surpasses the case when strange mesons are included, additionally, the strength of strange mesons will also slightly increase. In RMF theory ,the scalar meson $\sigma$ and $\sigma^*$ provides attraction, the vector meson $\omega$ and $\phi$ provides repulsion, these changes in meson strength may give the stiffer EOS.
\begin{table}
\caption{\label{tab:Table 4.}
Threshold densities $n_{cr}$ (in units of $\rho$/$\rho_0$) for $\Delta$ resonance in dense nuclear matter for different values of $x_{\sigma \Delta}$ without $\sigma\mbox{-}$cut scheme.}
\begin{ruledtabular}
\begin{tabular}{c|ccc}
\multirow{2}{*}{$\Delta$} & \multicolumn{3}{c}{$n_{cr}$ (without $\sigma^*$$\phi$)/(with $\sigma^*$$\phi$)} \\
 & \multicolumn{1}{c}{$x_{\sigma \Delta}=1.05$} & $x_{\sigma \Delta}=1.1$ & $x_{\sigma \Delta}=1.15$ \\ \hline
$\Delta^{++}$ & \multicolumn{1}{c}{/} & 7.54/7.73 & 6.27/6.24 \\
$\Delta^{+}$ & \multicolumn{1}{c}{8.60/8.83} & 6.53/6.60 & 5.01/5.04 \\
$\Delta^{0}$ & \multicolumn{1}{c}{6.53/6.66} & 4.49/4.56 & 3.52/3.55 \\
$\Delta^{-}$ & \multicolumn{1}{c}{2.13/2.13} & 1.90/1.90 & 1.74/1.74 \\
\end{tabular}
\end{ruledtabular}
\end{table}
\begin{figure}
\centering
\includegraphics{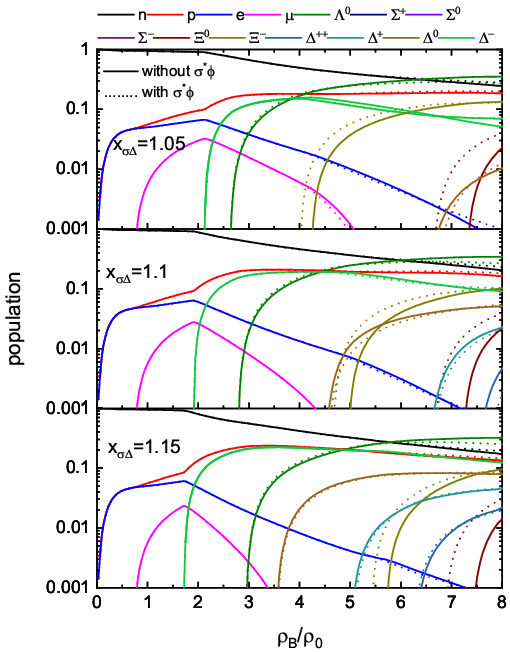}
\caption{Relative population of particles versus baryon density without $\sigma$-cut scheme with considering $\sigma^*$ and $\phi$ or not, $x_{\sigma \Delta}$=1.05, $x_{\sigma \Delta}$=1.1, $x_{\sigma \Delta}$=1.15.}
\label{fig4}
\end{figure}
\par

Fig. \ref{fig4} illustrates the relative population of particles as a function of baryon density for different values of $x_{\sigma \Delta}$, namely $x_{\sigma \Delta}=1.05, 1.1, 1.15$. The dashed and solid lines represent the cases with and without considering strange mesons ($\sigma^*$, $\phi$), respectively. When the strange mesons are taken into account, the critical densities of $\Xi^0$, $\Xi^-$, and $\Delta^{++}$ shift towards lower density regions, while the critical densities of $\Delta^+$ and $\Delta^0$ shift towards higher density regions. Additionally, with the increase of $x_{\sigma \Delta}$, the critical densities of $\Lambda^{0}$, $\Xi^{0}$, and $\Xi^{-}$ move towards higher density regions, whereas the critical densities of leptons move towards lower density regions. Notably, as $x_{\sigma \Delta}$ increases, the $\Delta^{++}$ resonance starts to appear, and the overall critical density for the $\Delta$ resonance decreases, pushing the appearance of hyperons to lower density regions. Regarding the critical density of $\Delta$ resonance we have placed in Table \ref{tab:Table 4.}.

\par
\begin{figure}
\centering
\includegraphics{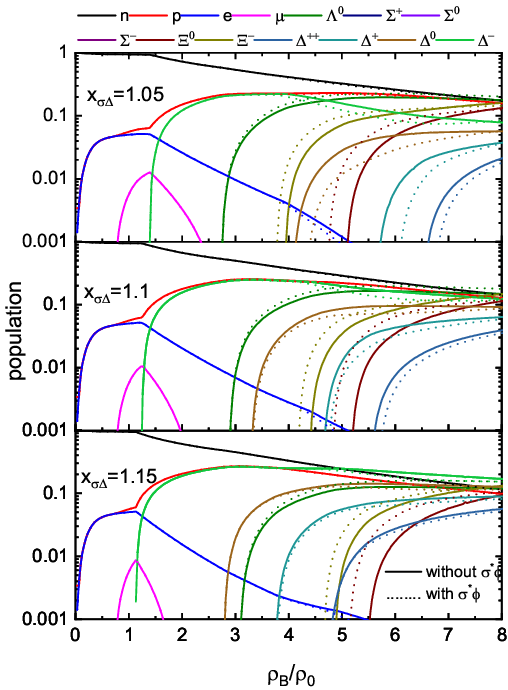}
\caption{Relative population of particles versus baryon density with considering $\sigma^*$ and $\phi$ or not, and $c_\sigma=0.15$, $x_{\sigma \Delta}$=1.05, $x_{\sigma \Delta}$=1.1, $x_{\sigma \Delta}$=1.15.}
\label{fig5}
\end{figure}

Next we examine the effect of the $\sigma$-cut scheme on the particle population, this is plotted in the Fig. \ref{fig5}. As the decrease of $\mu^{-}$, the $\Delta^{+}$ and $\Delta^{++}$ increase as the charge balance conditions lead to the increase of $\Xi^{-}$ and $\Delta^{-}$, suggesting that baryons are more favorable as neutralizers of positive charges compared to leptons. As $x_{\sigma \Delta}$ increases from 1.05 to 1.15, the critical value of $\Delta$ resonance shifts to lower density while the critical density of hyperons move toward the high density region, in particular. When $x_{\sigma \Delta}=1.15$, the critical density of $\Delta^{0}$ moves before the $\Lambda^{0}$. Although the $\sigma$-cut scheme significantly affects the critical density distribution of various particles, it does not change the relationship between the $\Delta$ resonance and strange mesons($\sigma^*$, $\phi$) as $x_{\sigma \Delta}$ varies.
\begin{figure}
\includegraphics{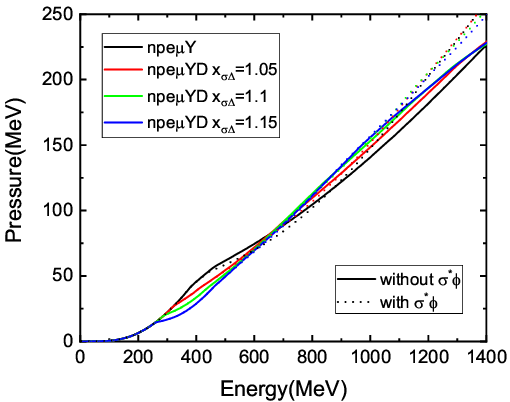}
\caption{Pressure  as a function of energy density without the $\sigma$-cut scheme. The black solid line is for n, p, leptons and hyperons whereas others are with additional $\Delta$ resonance, dotted lines contain $\sigma^*$ and $\phi$ mesons.}
\label{fig6}
\end{figure}
\par
Fig. \ref{fig6} shows the pressure as a function of energy density in neutron star matter containing $\Delta$ resonances without the $\sigma$-cut scheme. The dashed line represents the case when strange mesons $\sigma^*$ and $\phi$ are considered, while the solid line represents the case without considering strange mesons. Although not particularly significant in the low-energy density region, in the high-energy density region, the presence of strange mesons slightly stiffens the equation of state due to their attractive effect. As $x_{\omega\Delta}$ increases, the equation of state becomes softer in the energy density range from $300 MeV/fm^3$ to $600 MeV/fm^3$. However, for energy density greater than $600 MeV/fm^3$, it becomes significantly stiffer compared to the case when only hyperons are included. This suggests the existence of a softer equation of state in the low-density region, which ultimately constrains the radius of the neutron star, while the maximum mass does not show significant changes.

\begin{figure}
\centering
\includegraphics{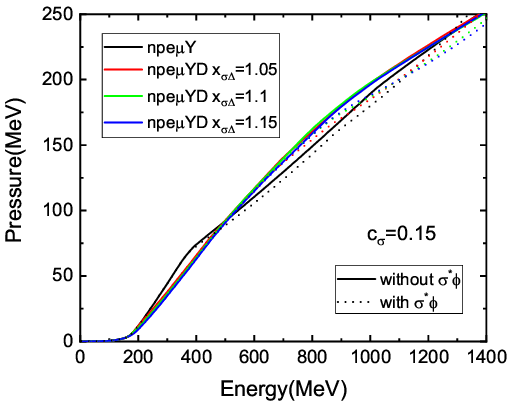}
\caption{Pressure as a function of energy density with the $\sigma$-cut scheme($c_\sigma=0.15$). The black solid line is for n, p, leptons and hyperons whereas others are with additional $\Delta$ resonance, dotted lines contain $\sigma^*$ and $\phi$ mesons.}
\label{fig7}
\end{figure}
When considering the $\sigma$-cut scheme, we plot the equation of state (EOS) in Fig. \ref{fig7}, where the dashed and solid lines represent the cases with and without strange mesons, respectively. We can observe that the $\sigma$-cut scheme significantly stiffens the EOS, and this is due to the truncation of the $\sigma$ meson field strength as shown in Fig. \ref{fig3}. Interestingly, in this case, the inclusion of strange mesons actually softens the EOS in the high-energy density region. Moreover, the $\sigma$-cut scheme retains the softening feature of the EOS in the low-density region. Compared to the case without the $\sigma$-cut scheme, the softening region shifts by approximately $50 MeV/fm^3$ towards the low-density region. By obtaining the EOS through this approach, we can solve the TOV equation to produce neutron stars with masses up to $2M_{\odot}$, effectively eliminating the "hyperon puzzle."
\begin{figure}
\centering
\includegraphics{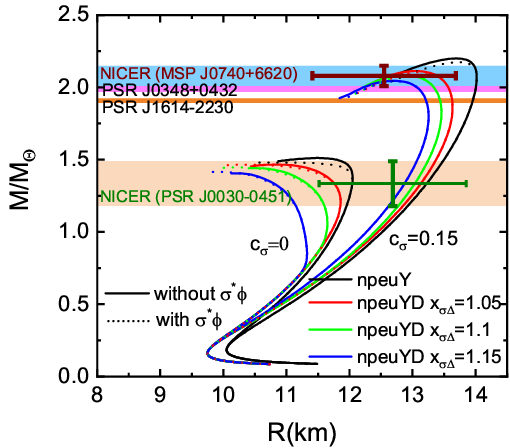}
\caption{Mass-radius relation using and not using $\sigma$-cut scheme in neutron star matter described by, the dotted line indicates that considering $\sigma^*$ and $\phi$. The horizontal bars indicate the observational constraints of PSR J1614 - 2230 \cite{Demorest:2010bx,Ozel:2010bz,NANOGrav:2017wvv,Fonseca:2016tux}, PSR J0348 + 0432 \cite{antoniadis2013massive}, MSP J0740 + 6620 \cite{fonseca2021refined} and PSR J0030-0451 \cite{riley2019nicer}.}
\label{fig8}
\end{figure}
\par
The mass-radius relationship for neutron stars (NS) discussed here and depicted in Fig. \ref{fig8}. The shaded bands represent the constraints imposed by the observables of massive neutron stars, namely PSR J1614-2230 \cite{Demorest:2010bx, Ozel:2010bz, NANOGrav:2017wvv, Fonseca:2016tux} and PSR J034+0432 \cite{antoniadis2013massive}. In 2019, the Neutron Star Interior Composition Explorer (NICER) collaboration reported precise measurements of the mass and radius of PSR J0030+0451 \cite{riley2019nicer}, and in 2021, they reported on MSP J0740+6620 \cite{fonseca2021refined}. The left solid lines in the figure, without $\sigma$-cut, demonstrate that different coupling parameters $x_{\sigma \Delta}$ have a notable impact on the maximum mass and radius of the neutron star. It reveals that the $\Delta$ resonance decreases the maximum mass and radius of the neutron star. As $x_{\sigma \Delta}$ increases (from 1.05 to 1.15), the maximum mass decreases. The right solid lines represent $c_{\sigma}$ = 0.15, which significantly boosts the maximum mass of the neutron star beyond $2M_\odot$, in agreement with the constraints from gravitational waves and NICER (MSP J0740+6620). However, there is no significant difference between the maximum mass and radius variation of neutron stars with the addition of $\sigma^*$ and $\phi$. Table \ref{Table5.} presents the simultaneous measurements of the radius for MSP J0740+6620 and PSR J0030-0451 using NICER data, along with the maximum mass of the neutron star for various values of $x_{\sigma \Delta}$.
\par
\begin{table*}[ht]
\caption{\label{Table5.}
The maximum mass (in unit of solar mass $M_{\odot}$), central energy density $\rho_c$ (in unit of $\rho$/$\rho_0$) and radius(km) in neutron star matter including strange mesons using and not using $\sigma$-cut scheme.}
\begin{ruledtabular}
\begin{tabular}{c|ccc|ccc|cc|cc}
\multirow{2}{*}{} & \multicolumn{2}{c}{without ${\sigma}$-cut} & & \multicolumn{2}{c}{ with ${\sigma}$-cut} & & \multicolumn{2}{c|}{MSP J0740+6620 \cite{fonseca2021refined}} & \multicolumn{2}{c}{PSR J0030-0451 \cite{riley2021nicer}} \\
 & \multicolumn{1}{c}{M} & ${\rho_c}$ & R & \multicolumn{1}{c}{M} & ${\rho_c}$ & R & \multicolumn{1}{c}{M} & R & \multicolumn{1}{c}{M} & R \\ \hline
($n,p,Y$) & \multicolumn{1}{c}{1.46} & 7.76 & 10.65 & \multicolumn{1}{c}{2.15} & 3.84 & 13.79 & \multicolumn{1}{c}{\multirow{5}{*}{$2.08\pm0.07$}} & \multirow{5}{*}{$12.39^{+1.3}_{-0.98}$} & \multicolumn{1}{c}{\multirow{5}{*}{$1.34^{+0.15}_{-0.16}$}} & \multirow{5}{*}{$12.71^{+1.14}_{-1.19}$} \\
$x_{\sigma \Delta}=1.05$($n,p,Y,D$) & \multicolumn{1}{c}{1.45} & 7.76 & 10.62 & \multicolumn{1}{c}{2.08} & 4.4 & 13.3 & \multicolumn{1}{c}{} &  & \multicolumn{1}{c}{} &  \\
$x_{\sigma \Delta}=1.1$($n,p,Y,D$) & \multicolumn{1}{c}{1.43} & 8.09 & 10.42 & \multicolumn{1}{c}{2.06} & 4.53 & 13.1 & \multicolumn{1}{c}{} &  & \multicolumn{1}{c}{} &  \\
$x_{\sigma \Delta}=1.15$($n,p,Y,D$) & \multicolumn{1}{c}{1.40} & 8.6 & 10.1 & \multicolumn{1}{c}{2.02} & 4.66 & 12.91 & \multicolumn{1}{c}{} &  & \multicolumn{1}{c}{} &  \\
\end{tabular}
\end{ruledtabular}
\end{table*}
\begin{figure}
\centering
\includegraphics{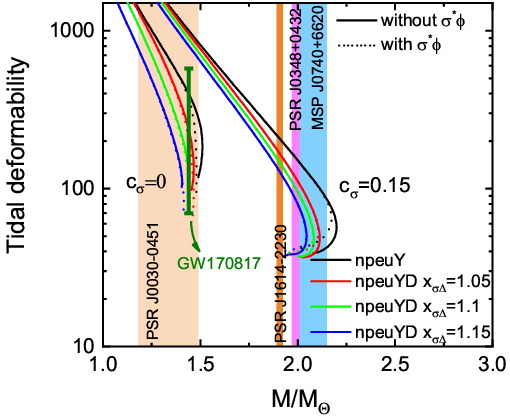}
\caption{The dimensionless $\Lambda$ as a function of star mass. The solid line indicates without $\sigma$-cut scheme, the dotted line indicates that considering $\sigma^*$ and $\phi$. And the constraints from GW170817 event for tidal deformability is shown.}
\label{fig9}
\end{figure}
\par
Another important constraint is the tidal deformability of the compact stars. In Fig. \ref{fig9} we present the tidal deformabilities of compact stars corresponding to those in Fig. \ref{fig8}. Based on the gravitational wave data from the binary neutron star merger event GW170817, the tidal deformability at 1.4$M_{\odot}$ was extracted as $\Lambda_{1.4}=190^{+390}_{-120}$ \cite{LIGOScientific:2018cki}. From the figure, it is evident that the $\sigma$-cut scheme with a stiffer equation of state (EOS) yields larger values of $\Lambda_{1.4}$ and heavier masses. However, these values of $\Lambda_{1.4}$ fall outside the constraint set by GW170817. On the other hand, the softer EOS, without the $\sigma$-cut scheme, satisfies the GW170817 constraint, resulting in smaller radii. Additionally, the inclusion of the $\Delta$ resonance maintains $\Lambda_{1.4}$ within the bounds of GW170817. These findings indicate that considering the $\Delta$ resonance in the softer EOS is necessary, given the strong constraint imposed by the observational tidal deformability of compact stars during the GW170817 event. Furthermore, future gravitational wave events from binary neutron star mergers are expected to provide measurements of the neutron star's tidal deformability at 2.0$M_{\odot}$.
\section{\label{sec:level4}Summary}
In this paper, we have discussed the $\Delta$ resonance and strange mesons ($\sigma^*, \phi$) within neutron stars under the IUFSU model, prompted by recent astronomical observations yielding rapid results on the radii and tidal deformations of compact stars. However, the maximum masses of neutron stars generated by the softer equation of state (EOS) (known as the hyperon puzzle) fail to approach 2.0$M_{\odot}$, thereby not satisfying the constraints from observations of massive neutron stars. Consequently, we employ the $\sigma$-cut scheme, resulting in a maximum mass exceeding 2$M_{\odot}$.
\par{}
We investigated the impact of strange mesons on neutron stars and found that within a neutron star, $\sigma^*$ and $\phi$ mesons shift the critical density of hyperons towards the low-density region. However, with the inclusion of the $\Delta$ resonance, strange mesons lead to a shift of the critical density of the $\Delta$ resonance towards the high-density region. As the coupling parameter $x_{\sigma \Delta}$ increases, the $\Delta$ resonance appears earlier, suggesting that the presence of strange mesons affects the critical density of the $\Delta$ resonance. These results indicate that although $\sigma^*$ and $\phi$ only interact with hyperons, considering the $\Delta$ resonance, they play a minor role in the interaction confinement between baryons. Additionally, the inclusion of strange mesons slightly increases the mass, but the variations are not significant. Interestingly, when the $\sigma$-cut scheme is considered, $\sigma^*$ and $\phi$ mesons lead to a softening of the equation of state.
\par{}
Furthermore, we explored the effect of $x_{\sigma \Delta}$ on the $\Delta$ resonance. For the $\Delta$ coupling constants, we consider $x_{\sigma \Delta}=1.05$, 1.1, and 1.15. The value of $x_{\sigma \Delta}$ significantly influences the relative population of particles as a function of the baryon density. We observe that the inclusion of the $\Delta$ resonance shifts the critical density of hyperons towards the high-density region as $x_{\sigma \Delta}$ increases from 1.05 to 1.15, while the critical density of the $\Delta$ resonance moves towards the low-density region, this suggests that an early appearance of the $\Delta$ resonance may contribute to the stability of neutron stars. Furthermore, with increasing $x_{\sigma \Delta}$, the equation of state softens in the low-density region, resulting in a significant reduction in the radius of neutron stars, while the maximum mass remains almost unchanged.
\par{}
When not using the $\sigma$-cut scheme, the softer equation of state considering the $\Delta$ resonance still falls within the $\Lambda_{1.4}$ range of GW170817 and results in smaller radii. And when we employ the $\sigma$-cut scheme with $c_{\sigma}=0.15$, we observe that the maximum mass and radius of neutron stars obtained align closely with the constraints from NICER (MSP J0740+6620). However, the tidal deformability exceeds the constraint from GW170817, for neutron stars with a mass exceeding $2M_{\odot}$, future gravitational wave events from binary neutron star mergers may provide new constraints on tidal deformability.

\bibliography{StrangeMeson}

\bibliographystyle{unsrt}

\end{document}